\documentclass[12pt]{article}
\usepackage{amsmath}
\usepackage{mathtools}
\usepackage{amssymb}
\usepackage{amsfonts}
\usepackage{graphicx}

\title{Cylindrical vector beam generator using a two-element interferometer}
\date{2019}

\begin{document}
\maketitle
\subsubsection*{Authors:}
Job Mendoza-Hern\'{a}ndez, Manuel F. Ferrer-Garcia, Jorge Arturo Rojas-Santana, and Dorilian Lopez-Mago$^{*}$
\subsubsection*{Addresses:}
Tecnologico de Monterrey, Escuela de Ingenier\'{i}a y Ciencias, Ave. Eugenio Garza Sada 2501, Monterrey, N.L., M\'{e}xico, 64849.\\
$^\ast$ dlopezmago@tec.mx

\section{abstract}
We realize a robust and compact cylindrical vector beam generator which consists of a simple two-element interferometer composed of a beam displacer and a cube beamsplitter. The interferometer operates on the higher-order Poincar\'{e} sphere transforming a homogeneously polarized vortex into a cylindrical vector (CV) beam. We experimentally demonstrate the transformation of a single vortex beam into all the well-known CV beams and show the operations on the higher-order Poincar\'{e} sphere according to the control parameters. Our method offers an alternative to the Pancharatnam-Berry phase optical elements and has the potential to be implemented as a monolithic device.

\section{Introduction}

Optical vector beams are structured  fields which posses a space-dependent polarization distribution \cite{Brown:10}. Two particular examples are the radially and azimuthally polarized beams, which are commonly called  cylindrical vector (CV) beams \cite{zhan_cylindrical_2009,Chen2018}. These are characterized by a donut shape intensity containing a central singularity surrounded by an azimuthally varying pattern of linear polarizations. The family of CV beams can be conveniently described as a coherent superposition of optical vortices with orthogonal polarizations and opposite helicities~\cite{Galvez:12}. The corresponding Jones vector of the transverse electric field can be written as
\begin{equation}
    \mathbf{E}(\mathbf{r})= \frac{1}{\sqrt{2}}\left[ \mathrm{LG}_{-m}(\mathbf{r})\, \mathbf{\hat{c}}_{R} + \exp(i\beta)\, \mathrm{LG}_{m}(\mathbf{r}) \, \mathbf{\hat{c}}_{L}\right].
    \label{EQ:CV}
\end{equation}
In the above equation, we have considered a monochromatic light beam, with wavelength $\lambda=2\pi/k$ and $\mathbf{r}=x\mathbf{\hat{x}}+y\mathbf{\hat{y}}$, propagating along the $z$-axis. The unit vectors $\mathbf{\hat{c}}_{R}$ and $\mathbf{\hat{c}}_{L}$ are the right- and left-handed unit polarization vectors, respectively. The function $\mathrm{LG}_{m}(\mathbf{r})$ represents an optical vortex whose complex amplitude is given by the single-ringed Laguerre-Gaussian (LG) beams. The LG beams belong to the class of helical modes with azimuthal dependence $\exp(im\phi)$, where the integer index $m$, also known as the topological charge, characterizes the helical wavefront of the beam. This helical wavefront indicates that the LG beams carry orbital angular momentum (OAM). In fact, for the LG beams, the amount of OAM per unit power and unit length is equal to $m$. For the particular case $m=1$, the phase difference $\beta$ in Eq.~(\ref{EQ:CV}) can be adjusted to generate radial, azimuthal or hybrid modes\cite{Wang:hybrid2010}. Higher-order polarization singularities can be realized with $m>1$ \cite{Otte:17,Freund}.

The CV beams have been extensively studied, in particular, due to their tight focusing properties which present a strong longitudinal component and a smaller focal spot compared to a focused Gaussian beam \cite{pu_tight_2010,porfirev_polarization_2016}.  They have been applied in multiple areas such as microscopy 
\cite{biss_dark-field_2006,hell_breaking_1994}, optical manipulation \cite{grier_revolution_2003,man_optical_2018,turpin_optical_2013,weng_creation_2014}, material processing \cite{kraus_microdrilling_2010, hnatovsky_role_2013,hamazaki_optical-vortex_2010,hnatovsky_polarization-dependent_2012}, and telecommunications \cite{gibson_free-space_2004,bozinovic_terabit-scale_2013}.

The generation of CV beams can be performed using active or passive methods\cite{zhan_cylindrical_2009}. In the active generation method the beams are obtained directly from the light source using a specially design intracavity resonator \cite{kozawa_generation_2005,ahmed_multilayer_2007}. The passive methods in free space are based on the modification of spatially-homogeneous polarized beams with devices that variant the polarization distribution. All the interferometric methods, where the wavefront is modified using spatial light modulators, belong to the passive methods \cite{rosales-guzman_simultaneous_2017,wang_generation_2007}. Other interferometric configurations have been explored which include compact and robust designs \cite{Chen2011,wang_generation_2007,li_efficient_2014,liu_compact_2019}. Furthermore, single-element vector beams generators have been introduced such as metasurfaces and Pancharatnam-Berry phase optical elements \cite{Capasso,Cardano:12}, which allow us to create complex polarization patterns based on a specific input beam. Nevertheless, these elements are not so common and their purchase or fabrication is not accessible for some laboratories.

The CV beams can be used as an extended basis to describe light beams with space-dependent polarization states, which are visualized in a higher-order Poincar\'{e} sphere (HOPS)  \cite{milione_higher-order_2011,PhysRevLett.108.190401,Holleczek2011}. In this new representation, the family of CV beams is described in a hybrid spatial-polarization basis, which is formed by the product of orthogonal spatial modes and circular polarization basis, i.e. the basis elements are $\{ \mathrm{LG}_{m} , \mathrm{LG}_{-m} \} \otimes \{ \mathbf{\hat{c}}_{R}, \mathbf{\hat{c}}_{L}\}$. This representation permits to easily visualize the effect of phase retarders and polarizers on the polarization pattern. Further details of this formalism can be found in Millione \textit{et al.} \cite{milione_higher-order_2011} and Holleczek \textit{et al.} \cite{Holleczek2011}. 

Due to the properties and applications of the CV beams, it is desirable to find alternatives to generate them. In this work, we introduce a compact and simple interferometric method to transform scalar vortex beams into CV beams using optical elements easily found in an optics laboratory. The main component of our method is a two-element interferometer composed of a beam displacer and a cube beam splitter with its semi-reflecting layer placed parallel to the optical axis of the system. We introduce the theoretical description for the CV beam generator and explain the effect of the control parameters on the higher-order Poincar\'{e} sphere. An experiment is performed to demonstrate our proposal by realizing all well-known CV beams.

\section{A two-element interferometer as a vector beam generator.}
\label{sec:CVBgen}

\begin{figure}[htbp]
\centering
\includegraphics[width=12 cm]{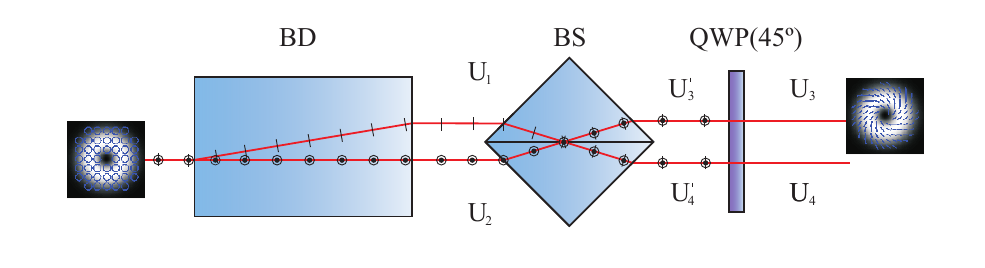}
\caption{Top view of the experimental scheme to generate CV beams. The input beam shown in the inset corresponds to an LG beam with homogeneous polarization. The LG beam passes through our two-element interferometer composed of a beam displacer (BD) and a cube beamsplitter (BS), Transverse lines and dots are used to indicate the horizontal and vertical polarization directions, respectively. The quarter-wave plate (QWP) is used to change the polarization basis. The output $\mathbf{U_{3}}$ is a CV beam as represented by the inset.}
\label{fig1}
\end{figure}

Our CV beam generator scheme is presented in Fig.~\ref{fig1}. It is mainly composed of a beam displacer (BD) and a cube beam splitter (CBS). Therefore, we call this device a \textit{two-element interferometer}. It is important to notice that the BD is not equivalent to a polarizing beam splitter. The BD separates the \textbf{s}- and \textbf{p}-polarization components of the input beam \textit{without} reflecting the \textbf{s} component. On the contrary, a polarizing beam splitter reflects the \textbf{s} component changing its helicity and hence its spatial mode. A requirement for our idea to work, it is that both beams arriving at the BS have the same spatial mode (except for an overall phase and amplitude factor). Therefore, the BD separates the beam according to our requirements. In addition, the output beams are parallel to the input beam, which facilitates alignment and reduces the number of optical elements. As explained later, a vector beam is already realized after the BS, however, in order to generate the polarization structure described by Eq.~(\ref{EQ:CV}), we use a quarter-wave plate (QWP) following the CBS to change the Cartesian polarization basis into a circular polarization basis. The input to the device is a vortex beam described as
\begin{equation}
    \mathbf{U_{in}}= \mathrm{LG}_{m} \, \mathbf{\hat{e}} = \mathrm{LG}_{m} (\cos\alpha \mathbf{\hat{x}}+ \exp{(i\theta)} \sin \alpha \mathbf{\hat{y}}),
    \label{eq1}
\end{equation}
where $\mathbf{\hat{x}},\mathbf{\hat{y}}$ are unit polarization vectors directed along the $x$ and $y$ directions, respectively. The unit vector $\mathbf{\hat{e}}= \cos\alpha \mathbf{\hat{x}}+ \exp{(i\theta)} \sin \alpha \mathbf{\hat{y}}$ is a generic elliptical polarization state with inclination angle $\alpha$ and constant phase difference $\theta$. The complex amplitude $\mathrm{LG}_{m}$ is given by 
\begin{equation}
\mathrm{LG}_{m}= C_m (r/w_0)^{|m|} \exp(-r^2/w_0^2)\exp(i m\phi),
\label{eq2}
\end{equation}
where we have used a cylindrical coordinate system $(x,y)=(r\cos\phi,r\sin\phi)$, $w_0$ is the beam waist at the plane $z=0$ and $C_m$ is a normalization constant, such that the beam has unit power, i.e. $\int \int |\mathrm{LG}_{m}|^{2} \mathrm{d}x \mathrm{d}y=1$. The optical vortex given by Eq.~\eqref{eq2} can be generated in multiple ways \cite{Yao:11}. In our case, taking advantage of a recent experiment, we opted to generate the vortex beam using a spatial light modulator (SLM) \cite{Arrizon:05}. Nevertheless, the vortex beam could be generated using a spiral phase plate or by transforming a Hermite-Gaussian mode using a pair of cylindrical lenses~\cite{Neil2000}. The vortex beam passes through a combination of a quarter-wave plate and a half-wave plate (not shown in Fig.~\ref{fig1}) in order to generate the elliptic polarization of Eq.~(\ref{eq1}). 

The vortex beam $\mathbf{U_{in}}$ enters the BD and is spatially separated in two orthogonally-polarized optical vortices $\mathbf{U_{1}}$ and $\mathbf{U_{2}}$ with the same topological charge and helicity. The outputs $\mathbf{U}_{1}$ and $\mathbf{U}_{2}$ after the BD can be written as
\begin{align}
        \mathbf{U_{1}}=& \cos\alpha \, \mathrm{LG}_{m} \, \mathbf{\hat{x}},\\
        \mathbf{U_{2}}=& \exp(i\theta)\sin \alpha \, \mathrm{LG}_{m} \, \mathbf{\hat{y}}.
\end{align}
In the above equation, we have neglected a phase difference between $\mathbf{U_{1}}$ and $\mathbf{U_{2}}$ that might arise due to the different optical paths between the ordinary ($\mathbf{U_{2}}$) and extraordinary modes ($\mathbf{U_{1}}$) of the BD. However, this phase difference can be compensated through phase $\theta$, which can be introduced using a quarter-wave plate. Nevertheless, in the experimental results that follow, we did not need to compensate for that phase difference.

After the BD, $\mathbf{U_1}$ and $\mathbf{U_2}$ are incident onto the CBS whose semi-reflecting layer is placed parallel to the propagation direction of the beams (refer to Fig.~\ref{fig1} to see the ray trajectories inside the CBS). This configuration was inspired by the single-element interferometer proposed by Ferrari \textit{et al.} \cite{ferrari_single-element_2007}. The CBS divides each input into two rays, one transmitted and one reflected. The transmitted parts of the beams are replicas of the input beams multiplied by a factor  $1/\sqrt{2}$. The reflected parts, in addition to the multiplicative factor $1/\sqrt{2}$, gain a phase-shift of $\pi/2$ and their OAM helicity (the sign of the topological charge $m$) is inverted. Notice that the trajectory of the reflected parts resembles the action of a Dove prism. It has been shown that an optical vortex inverts its OAM after traversing a Dove prism, as described by Gonzalez \textit{et al.} \cite{Gonzalez2006}. Therefore, for the reflected parts we make the transformation $m \rightarrow -m$. 

The output fields $\mathbf{U'_3}$ and $\mathbf{U'_4}$ after the CBS are given by the superposition of the transmitted and reflected parts as
\begin{align}
    \mathbf{U'_3}=& \frac{1}{\sqrt{2}}\left[ i\, \cos \alpha\,\, \mathrm{LG}_{-m} \, \mathbf{\hat{x}} \, +  \sin \alpha \exp(i\theta) \, \mathrm{LG}_{m} \, \mathbf{\hat{y}} \right], \\
   \mathbf{U'_4}=& \frac{1}{\sqrt{2}} \left[\cos \alpha\, \mathrm{LG}_{m} \, \mathbf{\hat{x}} \, + i \sin \alpha \exp(i\theta)  \, \mathrm{LG}_{-m} \, \mathbf{\hat{y}}\right].
\end{align}

Finally, in order to change the polarization basis from linear to circular, we use a quarter-wave plate whose fast axis is placed at 45 degrees with respect to the $x$-direction. The resulting beams are
\begin{align}
   \mathbf{U_3}=& \frac{i}{\sqrt{2}}\left[\, \cos \alpha \, \textrm{LG}_{-m} \, \mathbf{\hat{c}}_{R} \, -  \sin \alpha \exp(i\theta) \, \textrm{LG}_{m} \, \mathbf{\hat{c}}_{L}\right], \label{eq:out3} \\
    \mathbf{U_4}=&  \frac{1}{\sqrt{2}} \left[ \cos \alpha\, \textrm{LG}_{m} \, \mathbf{\hat{c}}_{R} \, +  \sin \alpha \exp(i\theta) \, \textrm{LG}_{-m} \, \mathbf{\hat{c}}_{L}\right], \label{eq:out4}
\end{align}
where $\mathbf{\hat{c}}_{R}= \mathbf{\hat{x}}-i \,
\mathbf{\hat{y}}$ and $\mathbf{\hat{c}}_{L}= \mathbf{\hat{x}}+i \,\mathbf{\hat{y}}$ are the right and left handed polarization vectors, respectively.  It must be noticed that the spatial distribution of the superposition can be modified only by varying the input beam polarization state. For example, the case with $\alpha=-\pi/4$ in Eq.~(\ref{eq:out3}) reproduces the polarization distribution given by Eq.~(\ref{EQ:CV}), which corresponds to the family of CV beams. 


\section{Experimental implementation}
The CV beam generator is constructed using a HeNe laser (Thorlabs HNL020LB) centered at a wavelength of $632.8$ nm. The single-ringed LG beams are generated applying a similar technique as the one used by Arrizon \textit{et al.} \cite{Arrizon:05} by using an amplitude-only SLM (HOLOEYE LC2002). The laser beam is spatially filtered, expanded and collimated before impinging the SLM. It is then passed through a combination of a 4f system with an aperture to recover the LG mode from the first diffraction order. Subsequently, the LG beam passes through the quarter- and half-wave plates to define the polarization state according to Eq.~(\ref{eq1}). 

The resulting LG beam goes through the BD (ThorLabs BD40) which provides a 4 mm spatial displacement between the two output beams. The output beams have a beam diameter of about 1 mm, and hence, the beam displacer separates the beam without overlapping them.

As mentioned in the previous section, both output beams are parallel to the input beam and have orthogonal polarizations. They enter the CBS which is aligned according to the scheme of Fig.~\ref{fig1}. The CBS is a 50:50 non-polarizing cube beamsplitter (Thorlabs BS013). 

Since the polarization distribution is constructed by the superposition of a reflected and transmitted beam, it must be noticed that the alignment of the CBS is crucial to ensure it. Following this condition, a three-axis mount has been used for the CBS in order to obtain a correct control over the superposition. After interfering at the beam splitter, the output beams are directed to a quarter-wave plate. The inhomogeneous polarization distribution is reconstructed by using traditional Stokes polarimetry
  \begin{eqnarray}
    S_0(x,y) &=& I_H (x,y)+ I_V (x,y) = I(x,y), \\
    S_1(x,y) &=& I_H (x,y) - I_V (x,y),\\
    S_2(x,y) &=& I_D (x,y)- I_A(x,y), \\
    S_3(x,y) &=& I_R(x,y)-I_L(x,y),
\end{eqnarray}
where $I(x,y)$ is the total transverse intensity of the beam and the subscripts $H,V,D,A,R,L$ are the horizontal, vertical, diagonal at $45^{\circ}$, diagonal at $135^{\circ}$, right and circular intensities, respectively. The intensity pattern of each output beam was captured using a CCD camera (Thorlabs DCU224M).

We compute the polarization ellipses at each spatial point of the transverse plane using the following equations \cite{Chipmanbook}:
\begin{eqnarray}
    \Psi = \frac{1}{2} \arctan \left( \frac{S_{2}}{S_{1}}\right),\label{eq:psi}\\
    \varepsilon = \frac{|S_{3}|}{\sqrt{S_{1}^{2}+S_{2}^{2}+S_{3}^{2}} + \sqrt{S_{1}^{2}+S_{2}^{2}}},\label{eq:ellip}
\end{eqnarray}
where $\Psi$ is the orientation of the major axis of the ellipse and $\varepsilon$ is the ellipticity. The handedness of the polarization state is determined by the sign of $S_3$. To illustrate the polarization distribution we consider a subset of equally spaced points on the transverse plane.

\section{Results and discussion}
\begin{figure}[htbp]
\centering
\includegraphics[width=12 cm]{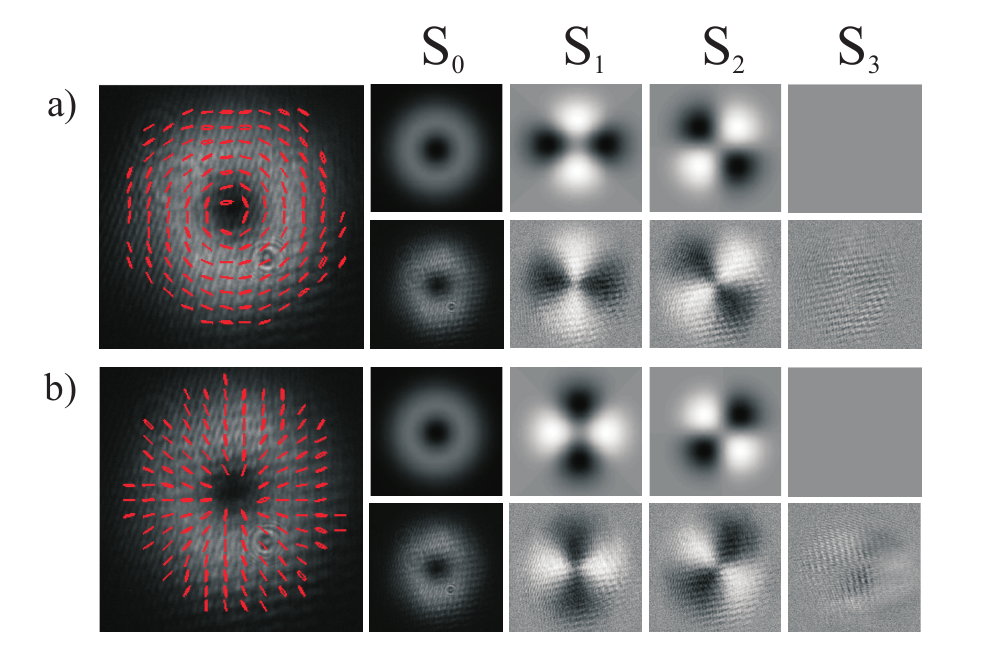}
\caption{Experimentally generated cylindrical vector beams. The results show the polarization distribution and their corresponding Stokes parameters. (a) Azimuthally-polarized CV beam produced with $m=-1$, $\alpha=\pi/4$, and  $\theta=0$ in Eq.~(\ref{eq:out3}), (b) Radially-polarized CV beam $(m=-1, \, \alpha=-\pi/4, \, \theta=0 )$. 
Theoretical (up) and experimental (down)  Stokes images.} 
\label{fig3:OR1} 
\end{figure}

We focus our experimental measurements on the output beam $\mathbf{U_3}$ given by Eq.~(\ref{eq:out3}). Notice that the output beam $\mathbf{U_4}$ provides similar polarization patterns. In fact, $\mathbf{U_4}$ is given by $\mathbf{U_3}$ with the transformations $m\rightarrow -m$ and $\theta \rightarrow \theta +\pi$. Figure \ref{fig3:OR1} shows our first experimental results. We realized the typical CV beams consisting of a radial and an azimuthal polarization pattern. Figure \ref{fig3:OR1}(a) shows the reconstructed polarization distribution along with the measured Stokes parameters for the azimuthally polarized mode. The intensity distribution $S_0$ is used as the intensity background for the polarization distribution, where the brighter regions represent larger intensities.  The polarization ellipses drawn on top of the intensity are computed with Eqs. \eqref{eq:psi} and \eqref{eq:ellip} employing the measured Stokes parameters shown next to the polarization distribution. The Stokes images are accompanied by theoretical simulations. The colormap for the Stokes images $S_1$, $S_2$ and $S_3$ represents positive values with the brighter regions and negative values with the darker regions.

Similarly, Fig.~\ref{fig3:OR1}(b) shows the results for the radially-polarized CV mode. In both results, the Stokes image $S_0$ shows the typical donut-shaped profile. However, the donut shape is not quite circular and presents astigmatism due to the CBS. In addition, the measured Stokes parameters show some background noise attributed to the waveplates and the protective IR window of the CCD camera. Furthermore, there is a small circular polarization component shown in the $S_3$ image, which should be ideally zero. These small deviations from the expected linear states on both cases are attributed to the angular dependence of the CBS, which introduces a small phase-shift that depends on the incidence angle \cite{Pezzaniti1994}. Nevertheless, we consider that the results show good agreement with the theoretical images and the polarization distribution clearly outlines the concentric rings and the radial lines characteristic of the azimuthally and the radially polarized modes, respectively.

\begin{figure}[htbp]
\centering
\includegraphics[width=12 cm]{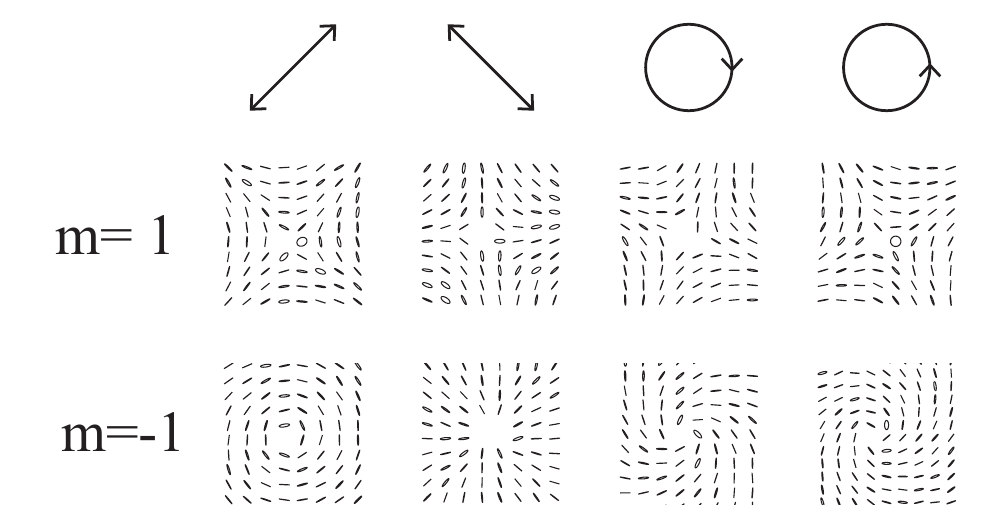}
\caption{Experimental measurement of the output polarization patterns according to the polarization (top row) and topological charge (left column) of the input beam $\mathbf{U_{in}}$.}
\label{fig4:comparative}
\end{figure}

Figure~\ref{fig4:comparative} shows further examples of the possible polarization distributions that can be generated with our method. For visualization purposes, we show the polarization ellipses without the background intensity. We generate the most common CV beams in terms of the topological charge and polarization state of the input field $\mathbf{U_{in}}$. Radial and azimuthal directions are both achieved when the input LG mode carries a topological charge of $m=-1$ and its polarization is in the diagonal basis, while the hybrid modes are obtained when $m=1$. Circular polarization in the input beam generates equally weighted superposition of the previous states, obtaining spiral polarization distributions when $m=-1$, and the hybrid modes when $m=1$. 

\begin{figure}[htbp]
\centering
\includegraphics[width=12 cm]{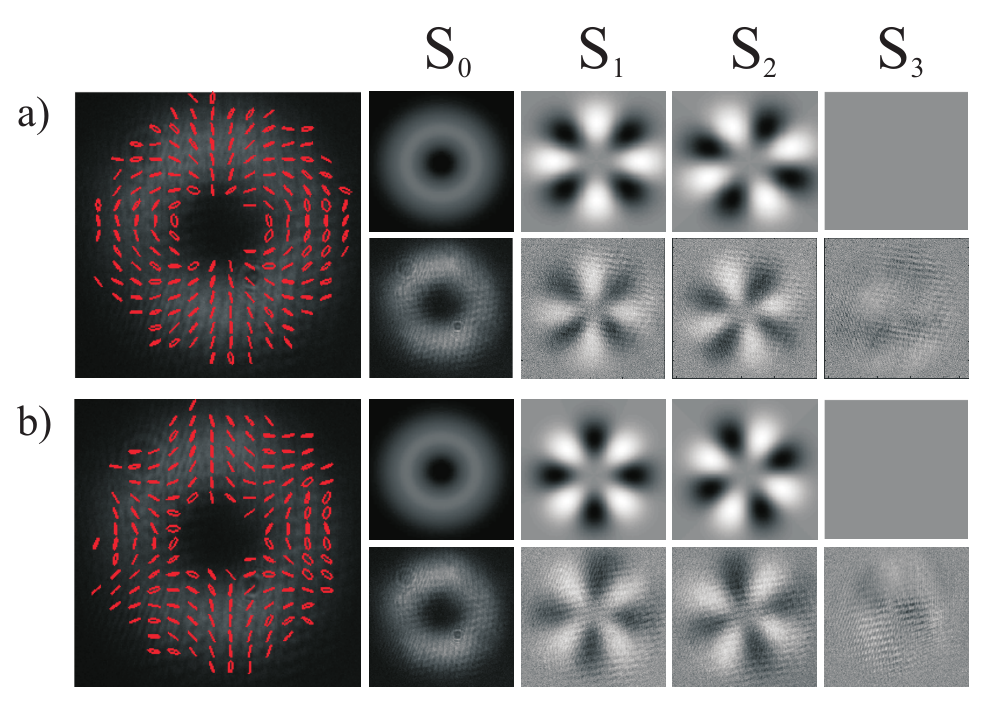}
\caption{Experimentally generated higher-order polarization singularities.  (a) a two-fold vectorial flower and (b) a six-fold vectorial spider web. Theoretical (up) and experimental (down)  Stokes images.}
\label{fig5:high}
\end{figure}

Finally, our proposal has the versatility of modifying the topological charge of the initial beam to obtain high-order polarization distributions\cite{Otte:17,Freund}. We obtained the polarization distributions, called flower and spider webs, as shown in Fig.~\ref{fig5:high}, where an LG beam with topological charge $m =\pm 2$ and antidiagonal polarization in the input field $\mathbf{U_{in}}$ is used. A polarization distribution of a flower is shown in Fig.~\ref{fig5:high}(a) and the spider web polarization distribution is shown in Fig.~\ref{fig5:high}(b).

\section{Transformations in the higher-order Poincar\'{e} sphere}
As shown in the above sections, the output polarization distribution is completely described by the polarization state and topological charge of the input beam. We can visualize the transformation of the input LG mode using the HOPS representation. A new set of Stokes parameters in the higher-order Poincar\'{e} sphere in terms of the input polarization parameters $(\alpha,\theta)$ are defined as \cite{milione_higher-order_2011}
 \begin{eqnarray}
    S'_1&=& \sin(2\alpha) \cos(\theta), \\
    S'_2&=& \sin(2\alpha) \sin(\theta), \\
    S'_3&=& \cos^2(\alpha)- \sin^2(\alpha).
\end{eqnarray}

\begin{figure}[htbp]
\centering
\includegraphics[width=\linewidth]{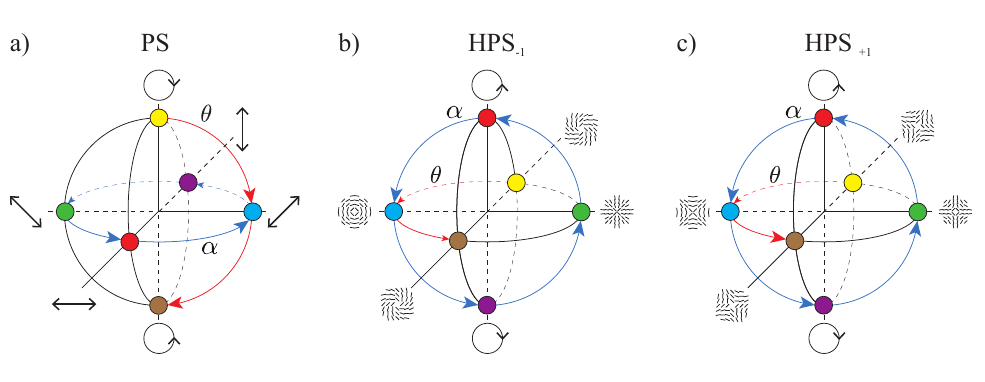}
\caption{Visual representation of the input and output polarization states and their transformations on the higher-order Poincar\'{e} sphere. Input and their respective output polarizations states are indicated by different color markers. The arrows indicate the increasing direction of the parameters in the intervals $\alpha=[0,2\pi]$ (red arrows) and $\theta=[-\pi/2,\pi/2]$ (blue arrows). (a) Input polarization state on the Poincar\'e sphere (PS). The corresponding output polarization distributions are represented on the higher-order Poincar\'{e} sphere (HPS) for (b) $m=1$ and (c) $m=-1$.}
\label{fig2:trans}
\end{figure}

 Figure \ref{fig2:trans} illustrates the input polarization state on the Poincar\'e sphere and its respective output polarization state on the corresponding HOPS, which depends on the topological charge of the input beam. From the new set of Stokes parameters, it is noticeable that a variation on the inclination angle $\alpha$ of the polarization state at the input beam represents a translation along a meridian. Meanwhile, a variation on the phase difference between the Cartesian components $\theta$ is mapped as a translation along the latitude. Both effects are illustrated in Figure \ref{fig2:trans} using blue and red arrows, respectively. Since the orthonormal basis of the HOPS depends on the sign of the topological charge, we obtain two independent spheres: one that includes the spirally polarized beams (in which the radial and azimuthal polarization directions are contained) when $m=-1$ and another one with the hybrid polarized beams when $m=1$. 

\section{Conclusions}
In this work, we have demonstrated the feasibility of a robust and compact cylindrical vector beam generator which consists of a simple two-element interferometer composed of a beam displacer and a cube beamsplitter. The interferometer operates on the higher-order Poincar\'{e} sphere transforming a homogeneously polarized vortex into a CV beam. We experimentally demonstrated the transformation of a single vortex beam into all the well-known CV beams and higher-order polarization singularities, and showed the operations on the higher-order Poincar\'{e} sphere according to the control parameters. Our method offers an alternative to the Pancharatnam-Berry phase optical elements and has the potential to be implemented in a monolithic device. Furthermore, since our method only employs refractive elements with a high-damage threshold, it can be used to create high-power vector beams.

\section*{Funding}
Consejo Nacional de Ciencia y Tecnolog\'{i}a (CONACYT) (Grants: 257517, 280181, 293471, 295239, APN2016-3140).

\section*{Acknowledgments}
JMH acknowledges partial support from CONACyT, M\'{e}xico. JMH thanks Israel Melendez Montoya for his help with the polarimeter.

\bibliographystyle{unsrt}
\bibliography{bibliography}

\begin{thebibliography}{10}

\bibitem{Brown:10}
Thomas~G. Brown and Qiwen Zhan.
\newblock Focus issue: Unconventional polarization states of light.
\newblock {\em Opt. Express}, 18:10775--10776, 2010.

\bibitem{zhan_cylindrical_2009}
Qiwen Zhan.
\newblock Cylindrical vector beams: from mathematical concepts to applications.
\newblock {\em Advances in Optics and Photonics}, 1:1--57, 2009.

\bibitem{Chen2018}
Jian Chen, Chenhao Wan, and Qiwen Zhan.
\newblock {Vectorial optical fields: recent advances and future prospects}.
\newblock {\em Science Bulletin}, 63:54--74, 2018.

\bibitem{Galvez:12}
Enrique~J. Galvez, Shreeya Khadka, William~H. Schubert, and Sean Nomoto.
\newblock {Poincar\'{e}-beam patterns produced by nonseparable superpositions
  of Laguerre--Gauss and polarization modes of light}.
\newblock {\em Appl. Opt.}, 51:2925--2934, 2012.

\bibitem{Wang:hybrid2010}
Xi-Lin Wang, Yongnan Li, Jing Chen, Cheng-Shan Guo, Jianping Ding, and Hui-Tian
  Wang.
\newblock A new type of vector fields with hybrid states of polarization.
\newblock {\em Opt. Express}, 18:10786--10795, 2010.

\bibitem{Otte:17}
Eileen Otte, Kemal Tekce, and Cornelia Denz.
\newblock Tailored intensity landscapes by tight focusing of singular vector
  beams.
\newblock {\em Opt. Express}, 25:20194--20201, 2017.

\bibitem{Freund}
Isaac Freund.
\newblock Polarization flowers.
\newblock {\em Opt. Comm.}, 199:47--63, 2001.

\bibitem{pu_tight_2010}
Jixiong Pu and Zhiming Zhang.
\newblock Tight focusing of spirally polarized vortex beams.
\newblock {\em Optics \& Laser Technology}, 42:186--191, 2010.

\bibitem{porfirev_polarization_2016}
Alexey~P. Porfirev, Andrey~V. Ustinov, and Svetlana~N. Khonina.
\newblock Polarization conversion when focusing cylindrically polarized vortex
  beams.
\newblock {\em Scientific Reports}, 6:6, 2016.

\bibitem{biss_dark-field_2006}
David~P. Biss, Kathleen~S. Youngworth, and Thomas~G. Brown.
\newblock Dark-field imaging with cylindrical-vector beams.
\newblock {\em Applied Optics}, 45:470--479, 2006.

\bibitem{hell_breaking_1994}
Stefan~W. Hell and Jan Wichmann.
\newblock Breaking the diffraction resolution limit by stimulated emission:
  stimulated-emission-depletion fluorescence microscopy.
\newblock {\em Optics Letters}, 19:780--782, 1994.

\bibitem{grier_revolution_2003}
David~G. Grier.
\newblock A revolution in optical manipulation.
\newblock {\em Nature}, 424:810, 2003.

\bibitem{man_optical_2018}
Zhongsheng Man, Zhidong Bai, Jinjian Li, Shuoshuo Zhang, Xiaoyu Li, Yuquan
  Zhang, Xiaolu Ge, and Shenggui Fu.
\newblock Optical cage generated by azimuthal- and radial-variant vector beams.
\newblock {\em Applied Optics}, 57:3592--3597, 2018.

\bibitem{turpin_optical_2013}
A.~Turpin, V.~Shvedov, C.~Hnatovsky, Yu~V. Loiko, J.~Mompart, and
  W.~Krolikowski.
\newblock Optical vault: A reconfigurable bottle beam based on conical
  refraction of light.
\newblock {\em Optics Express}, 21:26335--26340, 2013.

\bibitem{weng_creation_2014}
Xiaoyu Weng, Xiumin Gao, Hanming Guo, and Songlin Zhuang.
\newblock Creation of tunable multiple 3d dark spots with cylindrical vector
  beam.
\newblock {\em Applied Optics}, 53:2470--2476, 2014.

\bibitem{kraus_microdrilling_2010}
Martin Kraus, Marwan~Abdou Ahmed, Andreas Michalowski, Andreas Voss, Rudolf
  Weber, and Thomas Graf.
\newblock Microdrilling in steel using ultrashort pulsed laser beams with
  radial and azimuthal polarization.
\newblock {\em Optics Express}, 18:22305--22313, 2010.

\bibitem{hnatovsky_role_2013}
Cyril Hnatovsky, Vladlen~G. Shvedov, and Wieslaw Krolikowski.
\newblock The role of light-induced nanostructures in femtosecond laser
  micromachining with vector and scalar pulses.
\newblock {\em Optics Express}, 21:12651--12656, 2013.

\bibitem{hamazaki_optical-vortex_2010}
Junichi Hamazaki, Ryuji Morita, Keisuke Chujo, Yusuke Kobayashi, Satoshi Tanda,
  and Takashige Omatsu.
\newblock Optical-vortex laser ablation.
\newblock {\em Optics Express}, 18:2144--2151, 2010.

\bibitem{hnatovsky_polarization-dependent_2012}
Cyril Hnatovsky, Vladlen~G. Shvedov, Natalia Shostka, Andrei~V. Rode, and
  Wieslaw Krolikowski.
\newblock Polarization-dependent ablation of silicon using tightly focused
  femtosecond laser vortex pulses.
\newblock {\em Optics Lett.}, 37:226--228, 2012.

\bibitem{gibson_free-space_2004}
Graham Gibson, Johannes Courtial, Miles~J. Padgett, Mikhail Vasnetsov, Valeriy
  Pasâko, Stephen~M. Barnett, and Sonja Franke-Arnold.
\newblock Free-space information transfer using light beams carrying orbital
  angular momentum.
\newblock {\em Optics Express}, 12:5448--5456, 2004.

\bibitem{bozinovic_terabit-scale_2013}
Nenad Bozinovic, Yang Yue, Yongxiong Ren, Moshe Tur, Poul Kristensen, Hao
  Huang, Alan~E. Willner, and Siddharth Ramachandran.
\newblock Terabit-scale orbital angular momentum mode division multiplexing in
  fibers.
\newblock {\em Science}, 340:1545--1548, 2013.

\bibitem{kozawa_generation_2005}
Yuichi Kozawa and Shunichi Sato.
\newblock Generation of a radially polarized laser beam by use of a conical
  {Brewster} prism.
\newblock {\em Optics Letters}, 30:3063--3065, 2005.

\bibitem{ahmed_multilayer_2007}
Marwan~Abdou Ahmed, Andreas Voss, Moritz~M. Vogel, and Thomas Graf.
\newblock Multilayer polarizing grating mirror used for the generation of
  radial polarization in yb:yag thin-disk lasers.
\newblock {\em Optics Letters}, 32:3272--3274, 2007.

\bibitem{rosales-guzman_simultaneous_2017}
Carmelo Rosales-Guzm\'an, Nkosiphile Bhebhe, and Andrew Forbes.
\newblock {Simultaneous generation of multiple vector beams on a single SLM}.
\newblock {\em Optics Express}, 25:25697--25706, 2017.

\bibitem{wang_generation_2007}
Xi-Lin Wang, Jianping Ding, Wei-Jiang Ni, Cheng-Shan Guo, and Hui-Tian Wang.
\newblock Generation of arbitrary vector beams with a spatial light modulator
  and a common path interferometric arrangement.
\newblock {\em Optics Letters}, 32:3549--3551, 2007.

\bibitem{Chen2011}
Hao Chen, Jingjing Hao, Bai-Fu Zhang, Ji~Xu, Jianping Ding, and Hui-Tian Wang.
\newblock Generation of vector beam with space-variant distribution of both
  polarization and phase.
\newblock {\em Opt. Lett.}, 36:3179--81, 2011.

\bibitem{li_efficient_2014}
Si-Min Li, Sheng-Xia Qian, Ling-Jun Kong, Zhi-Cheng Ren, Yongnan Li, Chenghou
  Tu, and Hui-Tian Wang.
\newblock An efficient and robust scheme for controlling the states of
  polarization in a sagnac interferometric configuration.
\newblock {\em Europhysics Letters}, 105:64006, 2014.

\bibitem{liu_compact_2019}
Rui Liu, Ling-Jun Kong, Wen-Rong Qi, Shuang-Yin Huang, Zhou-Xiang Wang,
  Chenghou Tu, Yongnan Li, and Hui-Tian Wang.
\newblock Compact, robust, and high-efficiency generator of vector optical
  fields.
\newblock {\em Optics Letters}, 44:2382--2385, 2019.

\bibitem{Capasso}
J.~P. Balthasar~Mueller, Noah~A. Rubin, Robert~C. Devlin, Benedikt Groever, and
  Federico Capasso.
\newblock Metasurface polarization optics: Independent phase control of
  arbitrary orthogonal states of polarization.
\newblock {\em Phys. Rev. Lett.}, 118:113901, 2017.

\bibitem{Cardano:12}
Filippo Cardano, Ebrahim Karimi, Sergei Slussarenko, Lorenzo Marrucci, Corrado
  de~Lisio, and Enrico Santamato.
\newblock Polarization pattern of vector vortex beams generated by q-plates
  with different topological charges.
\newblock {\em Appl. Opt.}, 51:C1--C6, 2012.

\bibitem{milione_higher-order_2011}
Giovanni Milione, H.~I. Sztul, D.~A. Nolan, and R.~R. Alfano.
\newblock Higher-order poincar\'{e} sphere, stokes parameters, and the angular
  momentum of light.
\newblock {\em Physical Review Letters}, 107:053601, 2011.

\bibitem{PhysRevLett.108.190401}
Giovanni Milione, S.~Evans, D.~A. Nolan, and R.~R. Alfano.
\newblock Higher order pancharatnam-berry phase and the angular momentum of
  light.
\newblock {\em Phys. Rev. Lett.}, 108:190401, 2012.

\bibitem{Holleczek2011}
Annemarie Holleczek, Andrea Aiello, Christian Gabriel, Christoph Marquardt, and
  Gerd Leuchs.
\newblock {Classical and quantum properties of cylindrically polarized states
  of light}.
\newblock {\em Optics Express}, 19:9714, 2011.

\bibitem{Yao:11}
Alison~M. Yao and Miles~J. Padgett.
\newblock Orbital angular momentum: origins, behavior and applications.
\newblock {\em Adv. Opt. Photon.}, 3, 2011.

\bibitem{Arrizon:05}
Victor Arriz\'{o}n, Guadalupe M\'{e}ndez, and David~S\'{a}nchez de~La-Llave.
\newblock Accurate encoding of arbitrary complex fields with amplitude-only
  liquid crystal spatial light modulators.
\newblock {\em Opt. Express}, 13:7913--7927, 2005.

\bibitem{Neil2000}
Anna T~O Neil and Johannes Courtial.
\newblock {Mode transformations in terms of the constituent Hermite-Gaussian or
  Laguerre-Gaussian modes and the variable-phase mode converter}.
\newblock {\em Optics Communications}, 181:35--45, 2000.

\bibitem{ferrari_single-element_2007}
Jos\'{e}~A. Ferrari and Erna~M. Frins.
\newblock Single-element interferometer.
\newblock {\em Optics Communications}, 279:235--239, 2007.

\bibitem{Gonzalez2006}
N.~Gonz\'{a}lez, Gabriel Molina-Terriza, and Juan~P. Torres.
\newblock How a dove prism transforms the orbital angular momentum of a light
  beam.
\newblock {\em Opt. Express}, 14:9093, 2006.

\bibitem{Chipmanbook}
Young~G Chipman R~A, Lam W~T.
\newblock {\em Polarized light and optical systems}.
\newblock CRC Press, Boca Raton, FL, 2019.

\bibitem{Pezzaniti1994}
J.~Larry Pezzaniti and Russell~A. Chipman.
\newblock {Angular dependence of polarizing beam-splitter cubes}.
\newblock {\em Applied Optics}, 33:1916, 1994.

\end{thebibliography}

\end{document}